\documentclass[11pt]{article}
\usepackage{amsmath,amsthm,amssymb}
\usepackage{thmtools}
\usepackage{fullpage}
\usepackage[parfill]{parskip}
\usepackage[dvipsnames]{xcolor}
\usepackage[many]{tcolorbox}
\usepackage{libertine}
\usepackage{tikz}
\usepackage{hyperref}
\usepackage[nameinlink, capitalise]{cleveref}

\newtheorem{theorem}{Theorem}[section]
\newtheorem{lemma}[theorem]{Lemma}

\newtheorem{observation}[theorem]{Observation}
\theoremstyle{definition}
\newtheorem{definition}[theorem]{Definition}

\crefname{Program}{Program}{Programs}
\creflabelformat{Program}{(#2\textup{#1})#3}

\definecolor{linkc}{rgb}{0.1, 0.5, 0.7}
\definecolor{citec}{rgb}{0.6, 0.3, 0.7}
\definecolor{urlc}{rgb}{0.5, 0.1, 0.2}
\hypersetup{
    colorlinks=true,
    linkcolor=linkc,
    citecolor=citec,
    urlcolor=urlc
}

\newcommand{\multNM}[2]{#1-MNM-#2}
\newcommand{\multRule}{\textsc{NormalizedGeometricWinCountScore}}
\newcommand{\lambdaRule}{\textsc{NormalizedGeometricWinStrengthScore}}
\newcommand{\multTCCRule}
{\textsc{TopCycleGeometricWinCountScore}}

\newtcolorbox[auto counter,number within=section]{mybox}[2][]{colback=blue!8!white,colframe=gray!30!black,title={#2},#1}

\title{Strategyproof Tournament Rules for Teams with a Constant Degree of Selfishness}
\author{David Pennock \\ \normalsize DIMACS, Rutgers University \\ \normalsize\href{dpennock@dimacs.rutgers.edu}{dpennock@dimacs.rutgers.edu} \\
  \and Daniel Schoepflin \\ \normalsize DIMACS, Rutgers University \\ \normalsize\href{ds2196@dimacs.rutgers.edu}{ds2196@dimacs.rutgers.edu} \\
  \and Kangning Wang \\ \normalsize Rutgers University \\ \normalsize\href{kn.w@rutgers.edu}{kn.w@rutgers.edu} \\
}
\date{}

\begin{document}

\maketitle

\begin{abstract}
We revisit the well-studied problem of designing fair and manipulation-resistant tournament rules.  
In this problem, we seek a mechanism that (probabilistically) identifies the winner of a tournament after observing round-robin play among $n$ teams in a league. Such a mechanism should satisfy the natural properties of monotonicity and Condorcet consistency. Moreover, from the league's perspective, the winner-determination tournament rule should be \emph{strategyproof}, meaning that no team can do better by losing a game on purpose.

Past work considered settings in which each team is fully selfish, caring only about its own probability of winning, and settings in which each team is fully selfless, caring only about the total winning probability of itself and the team to which it deliberately loses. More recently, researchers considered a mixture of these two settings with a parameter $\lambda$. Intermediate selfishness $\lambda$ means that a team will not lose on purpose unless its pair gains at least $\lambda s$ winning probability, where $s$ is the individual team's sacrifice from its own winning probability. All of the dozens of previously known tournament rules require $\lambda = \Omega(n)$ to be strategyproof, and it has been an open problem to find such a rule with the smallest $\lambda$.

In this work, we make significant progress by designing a tournament rule that is strategyproof with $\lambda = 11$. Along the way, we propose a new notion of \emph{multiplicative} pairwise non-manipulability that ensures that two teams cannot manipulate the outcome of a game to increase the sum of their winning probabilities by more than a multiplicative factor $\delta$ and provide a rule which is multiplicatively pairwise non-manipulable for $\delta = 3.5$.
\end{abstract}

\section{Introduction}
We revisit the well-studied problem in computational social choice of designing (randomized) mechanisms to determine the winner of a tournament.  A \emph{tournament} of size $n$ is a complete directed graph on $n$ vertices.  The study of tournaments is foundational in social choice since they represent the pairwise majority graph among $n$ alternatives.  For example, the vertices could represent $n$ candidates for office with an edge from $i$ to $j$ indicating that a majority of voters prefer candidate $i$ to candidate $j$.  Alternatively, and perhaps more commonly, the tournament graph could represent the outcomes of the $n \choose 2$ matches in a round-robin sports tournament among $n$ teams in a league.  A tournament designer, then, aims to craft a \emph{tournament rule} $r$ which takes, as input, the tournament graph and outputs a winner distribution over the teams.  The designer's goal is to have the rule satisfy some desired fairness objective; that is, good teams should win more often than bad teams, but, since the vertices in the tournament graph correspond to self-interested agents, the designer must carefully craft her decision rule to avoid potential opportunities for strategic manipulations.  

To ensure fairness and non-manipulability, the two most ubiquitous and basic requirements for a tournament rule proposed in the literature are \emph{Condorcet consistency} and \emph{monotonicity}, respectively.  A rule $r$ is called Condorcet consistent if, whenever there is some team $i$ that beats all other teams, $r$ selects $i$ as the winning team with probability $1$.  A rule is called monotone if winning a match cannot decrease a team's chance of being selected (equivalently, no team can increase their probability of winning the tournament by intentionally throwing or losing a match that they can win). These requirements are minimal. Many rules, including trivial ones,\footnote{Take, for instance, the rule which outputs a Condorcet winner if one exists and otherwise choose a uniform random team.  This rule is clearly Condorcet consistent and it is also monotone, since winning a match weakly increases a team's winning probability.}
satisfy these notions. Moreover, monotonicity does not truly capture the breadth of strategic manipulations observed in practice.

In addition to individual manipulations within tournaments, there have been documented instances of teams \emph{colluding} in order to improve outcomes for their coalition.  One particularly infamous example is the so-called ``disgrace of Gij\'{o}n'' during the group stage of the 1982 FIFA World Cup, wherein  it appeared to many outside observers that the team from Austria and the team from West Germany fixed the outcome of the match (with West Germany defeating Austria by a $1$--$0$ margin) to ensure that both teams would proceed past the group stage to the knockout round \cite{shpigel2023important}.  Because of such manipulations, prior work in the literature has proposed the study of tournament rules which are \emph{pairwise} non-manipulable or robust against collusion between two teams.  As it turns out, the degree to which a tournament designer can prevent such pairwise manipulations is related to how selfish the teams are.  On one extreme, if two teams would collude only when \emph{neither} team's probability decreases (i.e., teams are completely selfish and would fix the match only if it would lead to a Pareto improvement in winning probabilities for the two) then there are rules which are Condorcet consistent and robust to pairwise collusion \cite{altman2010nonmanipulable2}.  On the other hand, if two teams care only about their shared probability of winning (i.e., teams are completely selfless and would fix the match as long as it increased the sum of their winning probabilities) then achieving Condorcet consistency and pairwise non-manipulability is impossible \cite{altman2010nonmanipulable1}.

Aiming to interpolate between these two extreme cases, Pennock et al. \cite{Pennock2024} defined a notion of parameterized selfishness which they call \emph{$2$-non-manipulable for $\lambda$} ($2$-NM$_\lambda$). A selfishness of $\lambda$ means that a team $i$ will not sacrifice $x$ probability by colluding with team $j$ (by throwing the match to $j$ that $i$ would otherwise win) unless $j$'s winning probability increases by $(\lambda + 1)x$.  Hence, a team with $\lambda$ selfishness 
will only collude and sacrifice some winning probability if their group's total winning probability increases by $\lambda$ times as much.  In particular, $\lambda = \infty$ captures complete selfishness and $\lambda = 0$ captures complete selflessness.  With their notion of $2$-NM$_\lambda$ in hand, \cite{Pennock2024} conjectured that $\lambda = 1$ is a sharp cutoff.  They proved that no rule is Condorcet consistent and $2$-NM$_\lambda$ for $\lambda < 1$ and demonstrated empirically that a Condorcet consistent and $2$-NM$_1$ rule exists for tournaments with at most six teams.  Unfortunately, their experimental approach did not scale effectively to larger tournaments and did not suggest any obvious general rule. Moreover, they demonstrate that all known rules require $\lambda = \Omega(n)$ to satisfy $2$-NM$_\lambda$ (or fail to satisfy it for any bounded $\lambda$), see \cite[Table 2]{Pennock2024}. So the gap in understanding that they left was surprising and large: even though no known rule did better than $\Omega(n)$, they suspected that $O(1)$---in fact precisely $1$---was possible. Notably, even some trivial tournament rules satisfy $2$-NM$_{\Theta(n)}$, meaning that the state of the art guarantees were very weak.\footnote{Consider the rule which outputs a Condorcet winner if one exists and otherwise chooses a team uniformly at random.  This rule satisfies $2$-NM$_{n-1}$ 
since each of the $n$ teams gets $1/n$ probability whenever there is no Condorcet winner.  As such, if changing the outcome of a game between $i$ and $j$ turns $j$ into a Condorcet winner, then $j$'s probability increases by $\frac{n-1}{n}$ while $i$'s probability decreases by $1/n$.} As such, they left open the following interesting question which forms the basis of our work:

\begin{quote}
    \textbf{Main Question: } \textit{What is the smallest possible $\lambda$ such that there exists a tournament rule satisfying Condorcet consistency and $2$-NM$_\lambda$?} 
\end{quote}

\subsection{Our Results}
In this paper, we make significant progress toward the question above by identifying a tournament rule \lambdaRule\  
which satisfies Condorcet consistency, monotonicity, and $2$-NM$_{11}$ for \emph{any} number of teams.  As discussed above, this is the first explicit tournament rule for tournaments with $n$ teams that guarantees $2$-NM$_\lambda$ for $\lambda = o(n)$.  

En route to this result, we first propose a novel notion of non-manipulability which we call \emph{$\delta$-multiplicative $k$-non-manipulability} (\multNM{$k$}{$\delta$}), which we believe may be of independent interest.  Loosely, a tournament rule 
satisfies \multNM{$k$}{$\delta$} if no coalition of $k$ teams can fix the outcomes of their matches in order to raise their joint winning probability by more than a \emph{multiplicative} factor $\delta$.  We then provide a  tournament rule, called \multRule, 
which achieves \multNM{$2$}{$3.5$} as well as monotonicity and Condorcet consistency, meaning that no two teams can collude and improve the sum of their winning probabilities by more than $3.5$ times.  \multRule\  achieves \multNM{$2$}{$3.5$} by carefully controlling the amount that any team's winning probability can increase or decrease by changing the outcome of a single match.  First, our rule ensures that any team $i$ which is not an \emph{almost-Condorcet winner} (i.e., wins strictly fewer than $n-2$ games) has a winning probability that can only increase by a small multiplicative factor by changing the result of any match involving $i$.  Second, our rule ensures that almost-Condorcet winners receive a large probability of winning.  To achieve these dual objectives, we define a scoring rule which maps the number of victories each team has to a numerical score.  Then \multRule\ selects each team $i$ to be the winner in proportion to the score of $i$ over the sum of all scores.  By choosing our scoring rule to be (roughly) a geometric sequence, we achieve our desired objectives and we prove that \multRule\ satisfies \multNM{$2$}{$3.5$}.

While the approach in \multRule\ appears close to achieving $2$-NM$_\lambda$, it suffers from a fatal flaw observed by Pennock et al. \cite{Pennock2024} common to many tournament rules.  In particular, for so-called \emph{superman-kryptonite graphs}---tournaments where a distinguished vertex $i$ called superman defeats all other teams except one distinguished vertex $j$ called kryptonite and all teams except $i$ defeat $j$---\multRule\ fails to give $2$-NM$_{o(n)}$.  For many existing tournament rules (including \multRule), this is because the kryptonite vertex receives a vanishing probability of winning while the superman vertex receives winning probability bounded away from $1$. This allows kryptonite to sacrifice $O(1/n)$ winning probability to give superman an $\Omega(1)$-probability gain, violating $2$-NM$_\lambda$ for $\lambda = o(n)$.  In other words, a crucial barrier for many rules is that they do not account for the possibility that a team with a very small number of wins could beat a team with a much larger number of wins. 
For our purposes, the key problem is that the scoring rule controlling the teams' winning probabilities in \multRule\ cares only about the \emph{number} of victories of each team rather than the \emph{strength of teams} each team defeats.
This observation leads to our second tournament rule \lambdaRule, which corrects this issue and achieves $2$-NM$_{11}$.    To do so, \lambdaRule\ defines a more complex scoring rule that rewards both the number of wins and the number of wins of their defeated teams, ensuring that the scores of any two teams $i$ and $j$ are related 
and depend on whether $i$ defeats $j$ or vice-versa.  With this, we can directly tie the gain in probability $j$ were to enjoy to the loss in probability that $i$ would suffer if $i$ threw the game to $j$, which allows us to guarantee $2$-NM$_{11}$. 

\section{Related Work}
Tournament design has been extensively studied in a wide range of disciplines, including computer science, mathematics, political science, and psychology. We cannot adequately catalog all of the works in this field herein and, instead, focus on the most relevant topics and direct the reader to \cite{brandt2016handbook} and \cite{laslier1997tournament}
for a survey of the field more generally.  

The model of competing strategic teams which we study was initially proposed in \cite{altman2010nonmanipulable1}, which studied deterministic rules, and \cite{altman2010nonmanipulable2}, which extended the framework to randomized tournament rules.  Altman et al. \cite{altman2010nonmanipulable1} demonstrated that no Condorcet consistent tournament rule is pairwise ``strongly non-manipulable'' ($2$-SNM), i.e., no rule can prevent a pair of teams from colluding to increase the sum of their winning probabilities.  On the other hand, Altman and Kleinberg \cite{altman2010nonmanipulable2} provided monotone, Condorcet consistent rules which are pairwise ``Pareto non-manipulable'', i.e., there exist rules which can prevent a pair of teams from colluding to increase both of their probabilities.  In the language of \cite{Pennock2024}, pairwise strong non-manipulability is exactly $2$-NM$_0$ and pairwise Pareto non-manipulability is $2$-NM$_\infty$.

Given the impossibility result of \cite{altman2010nonmanipulable1}, subsequent work considered relaxing the objective to approximate guarantees.  While the initial work of \cite{altman2010nonmanipulable1,altman2010nonmanipulable2} considered relaxing the fairness constraint (i.e., Condorcet consistency), a fruitful line of literature has emerged around relaxing the strategic constraint (i.e., $2$-SNM).  In particular, Schneider et al. \cite{schneider2017condorcet} defined the notion of ``$k$-wise strong non-manipulability up to $\alpha$'' ($k$-SNM-$\alpha$).  A rule satisfies $k$-SNM-$\alpha$ if and only if no group of up to $k$ teams can collude to increase the sum of their winning probabilities by an \emph{additive} $\alpha$.  They show that no Condorcet consistent rule satisfies $2$-SNM-$\alpha$ for any $\alpha < 1/3$ and also show that a rule they call \emph{Randomized Single Elimination Bracket} satisfies Condorcet consistency, monotonicity, and $2$-SNM-$1/3$.  Further work of \cite{schvartzman2020approximately} and \cite{ding2021approximately} proposed two alternative tournament rules called \emph{Randomized King of the Hill} and \emph{Randomized Death Match}, respectively, and showed that these rules achieve Condorcet consistency, monotonicity, and $2$-SNM-$1/3$, as well.  We note that while, at its surface, $k$-SNM-$\alpha$ appears closely related to our proposed notion of $k$-MNM-$\delta$, the required approaches to achieve guarantees seem different.  Indeed, as we show in \cref{obs:rseb-no-guar}, while the Randomized Single Elimination Bracket rule achieves the best possible guarantee of $2$-SNM-$1/3$, it fails to achieve $2$-MNM-$\delta$ for any bounded $\delta$.

Moving beyond pairwise manipulations, much less is known.  Schvartzman et al. \cite{schvartzman2020approximately} provided a non-monotone (but still Condorcet consistent) tournament rule which satisfies $k$-SNM-$2/3$ for any $k$, and they showed that no rule exists that satisfies Condorcet consistency and $939$-SNM-$1/2$.  Dinev and Weinberg \cite{dinev2022} later showed that Randomized Death Match achieves $3$-SNM-$31/60$ and Mik\v{s}an\'{\i}k et al. \cite{miksanik2024} provided a monotone and Condorcet consistent rule achieving both $2$-SNM-$1/3$ and $3$-SNM-$1/2$.

A parallel direction of work in computational tournament design examines how a designer can attempt to fix the outcomes of the random process in a tournament rule---typically the Randomized Single Elimination Bracket rule---to ensure that a certain team wins.  There is a significant line of literature in this domain, particularly focusing on the computational complexity of this problem, (see, e.g., \cite{kim2015fixing,aziz2018fixing}) which we do not attempt to catalog here.  We direct the interested reader to the survey of Suksompong \cite{suksompong2021tournaments} for further details. 
In our work, by contrast, we care only about the \emph{existence} of a tournament rule that satisfies $2$-NM$_\lambda$ for $\lambda = O(1)$. 
However, we note that our proposed tournament rules are computationally efficient and can easily be implemented with running time quadratic in the number of vertices in the tournament graph.

\section{Preliminaries}\label{sec:prelims}
We consider settings in which a central organizer is running a tournament between $n$ self-interested teams.

\begin{definition}[tournament]
    A (round robin) \emph{tournament} $T$ on a set $N = \{1, 2, \ldots, n\}$ of $n$ teams is the result of the $n \choose 2$ possible pairwise matches between those $n$ teams. It will frequently be useful to encode a tournament $T$ using a directed \emph{tournament graph} $G$ in which each vertex $i \in N$ denotes one of the $n$ teams and an edge from $i$ to $j$ in $G$ encodes that $i$ defeats $j$ in their pairwise match. We will use $\mathcal{T}_n$ to denote the set of all possible tournaments on $n$ teams.
\end{definition}

At several points in our presentation, we will need to consider two tournaments that have the same pairwise outcomes except for certain matches. For this purpose, we define the concept of neighboring tournaments.

\begin{definition}[$S$-adjacency]
    For a subset of teams $S \in N$, two tournaments $T, T' \in \mathcal{T}_n$ are called \emph{$S$-adjacent} if $T$ and $T'$ are identical except possibly for matches played between teams in $S$. In other words, an edge $(i, j)$ must have the same direction in the two tournament graphs of $T$ and $T'$ that are $S$-adjacent, as long as at least one of the endpoints $i$ and $j$ is not in $S$.
\end{definition}

\begin{definition}[tournament rule]
    A \emph{tournament rule} $r \colon \mathcal{T}_n \rightarrow \Delta^n$ (where $\Delta^n$ denotes the $n$-simplex) maps tournaments to probability distributions on the teams. The probability distribution $r(T)$ then encodes the probability that each team wins the tournament $T$, and we will often use $r_i(T)$ to denote the probability that the team $i$ wins the tournament $T$ under the rule $r$.
\end{definition}

Our organizer aims to design a tournament rule that satisfies certain fairness and competitiveness properties. A basic and natural property is Condorcet consistency, which stipulates that a team that wins all of its pairwise matches must win the tournament with certainty.

\begin{definition}[Condorcet consistency]
    In a tournament $T$, a team $i$ is called a \emph{Condorcet winner} if $i$ wins all the $n-1$ matches in which it participates. A tournament rule $r$ is called \emph{Condorcet consistent} if $r_i(T) = 1$ for every tournament $T$ and every team $i$ that is a Condorcet winner in $T$.
\end{definition}

To ensure competitiveness, a tournament rule should prevent teams from intentionally losing matches. In other words, a tournament rule should ensure that the winning probability of any team only increases as it wins more matches.

\begin{definition}[monotonicity]
    A tournament rule $r$ is called \emph{monotone} if for any $\{i, j\}$-adjacent tournaments $T$ and $T'$ in which $i$ beats $j$ in $T$ but not in $T'$, it holds that $r_i(T) \geq r_i(T')$.
\end{definition}

The monotonicity property is natural and corresponds to the feature of ``individual strategyproofness'' in tournament rules. However, a tournament rule should ideally go beyond individual strategyproofness and also guarantee a form of group strategyproofness, so that it can be robust to potential collusion between teams.

In particular, we hope to prevent all pairs of teams from fixing the outcome of their match for joint benefits. Perhaps a team $i$ would be willing to sacrifice some winning probability provided a co-conspirator team $j$ gained significantly in probability; in such cases, the amount of probability that $j$ would need to gain for $i$ to be willing to collude is captured by the ``selfishness'' of $i$. The following notion of \cite{Pennock2024} formalizes this idea; we state their definition and explain it below.

\begin{definition}[$2$-NM$_\lambda$ \cite{Pennock2024}]
    A tournament rule $r$ is called \emph{$2$-non-manipulable for $\lambda \geq 0$} (abbreviated as $2$-NM$_\lambda$) if $r_i(T') + r_j(T') \leq r_i(T) + r_j(T) + \lambda \max\{r_i(T) - r_i(T'), r_j(T) - r_j(T')\}$ for all teams $i$ and $j$ and all $\{i,j\}$-adjacent tournaments $T \neq T' \in \mathcal{T}_n$.
\end{definition}

Observe that if a tournament rule $r$ satisfies monotonicity and $\lambda > 0$ then the definition of $2$-NM$_\lambda$ simplifies to $r_j(T') - r_j(T) \leq (\lambda+1)(r_i(T) - r_i(T'))$ for any two $\{i,j\}$-adjacent tournaments $T \neq T'$ where $i$ defeats $j$ in $T$.  In other words, the loss in probability $x$ that team $i$ suffers moving from $T$ to $T'$ is compensated by no more than $(\lambda+1)x$ gain in probability for $j$ (and, hence, the coalition's winning probability increases by no more than $\lambda$ times the loss that $i$ suffers).  The notion of $2$-NM$_\lambda$ interpolates between two previously studied notions of pairwise manipulation called $2$-strong non-manipulability which corresponds to the equation obtained when $\lambda = 0$ and $2$-Pareto non-manipulability which corresponds to the equation obtained as $\lambda \rightarrow \infty$.  On one hand, tournament rules exist which guarantee monotonicity, Condorcet consistency, and $2$-Pareto non-manipulability \cite{altman2010nonmanipulable2}.  On the other hand, Condorcet consistency and $2$-strong non-manipulability turn out to be incompatible \cite{altman2010nonmanipulable1}.  In fact, $2$-NM$_\lambda$ and Condorcet consistency are incompatible for $\lambda < 1$, but all Condorcet consistent tournament rules in the literature fail to guarantee $2$-NM$_\lambda$ for any $\lambda = o(n)$.

In this work, we also propose a novel notion of groupwise non-manipulability which relates to the total multiplicative gain in probability a coalition of teams $S$ can procure via fixing the matches between them.  Designing tournament rules to ensure our new notion of  multiplicative non-manipulability for pairs of teams turns out to be a key stepping stone toward our main result of providing a Condorcet consistent and monotone tournament rule which satisfies $2$-NM$_\lambda$ for $\lambda = O(1)$.  

\begin{definition}[$\delta$-multiplicative $k$-non-manipulability]
    A tournament rule $r$ is \emph{$\delta$-multiplicative $k$-non-manipulable} for $\delta \geq 1$ (abbreviated as $k$-MNM-$\delta$) if for all subsets $S$ of at most $k$ players and all pairs of $S$-adjacent tournaments $T, T' \in \mathcal{T}_n$ we have that $\sum_{i \in S}r_i(T) \leq \delta\cdot \sum_{i \in S}r_i(T')$.  A tournament rule $r$ is $\delta$-multiplicative $2$-non-manipulable if for all pairs of teams $i,j$ and all $\{i,j\}$-adjacent tournaments $T, T' \in \mathcal{T}_n$ we have that $r_i(T) + r_j(T) \leq \delta(r_i(T') + r_j(T'))$.
\end{definition}

\section{Multiplicative Non-Manipulability}\label{sec:mnm-main}
We first consider pairwise non-manipulability in the form of $\delta$-multiplicative $2$-non-manipulability.  Recall that a tournament rule $r$ satisfies \multNM{$2$}{$\delta$} if no pair of teams can increase the sum of their winning probabilities by more than a $\delta$ multiplicative factor as a result of fixing the outcome of their match.  In other words, this notion is a multiplicative relaxation of $2$-strong non-manipulability (i.e., $2$-strong manipulability corresponds to \multNM{$2$}{$1$}) and, hence, is the multiplicative analogue of the notion of \emph{$2$-strong non-manipulability at probability $\alpha$} ($2$-SNM-$\alpha$) \cite{schneider2017condorcet} which has been extensively studied in the literature.\footnote{A tournament rule $r$ is $2$-strong non-manipulable at probability $\alpha$ if for all pairs of teams $i,j$ and all $\{i,j\}$-adjacent tournaments $T, T' \in \mathcal{T}_n$ we have that $r_i(T) + r_j(T) \leq r_i(T') + r_j(T') + \alpha$.}  We believe that \multNM{$2$}{$\delta$} is an interesting and natural notion in its own right, particularly given that multiplicative approximation is arguably a more standard goal in the broader computer science literature.  Notably, \multNM{$2$}{$\delta$} is often a more binding constraint than $2$-SNM-$\alpha$ for teams with \emph{small} winning probabilities.

From its relation to strong non-manipulability and $2$-SNM-$\alpha$, we can immediately obtain an impossibility for \multNM{$2$}{$\delta$}.  In particular, we can show that no Condorcet consistent rule is capable of guaranteeing better than \multNM{$2$}{$1.5$} from very similar reasoning as the proof in \cite{schneider2017condorcet} that achieving Condorcet consistency and better than $2$-SNM-$1/3$ is impossible.

\begin{observation}\label{obs:mult-nm-lb}
    There is no Condorcet consistent tournament rule on $n \geq 3$ players that is \multNM{$2$}{$(3/2 - \varepsilon)$} for any constant $\varepsilon > 0$.  
\end{observation}
\begin{proof}
    Suppose some rule $r$ guarantees Condorcet consistency and \multNM{$2$}{$(3/2 - \varepsilon)$} for $\varepsilon > 0$.  Consider the tournament $T$ with three teams $i$, $j$, and $k$ whose tournament graph forms a directed cycle (i.e., $i$ beats $j$ beats $k$ beats $i$).  Observe then, that any of the three teams could become a Condorcet winner as a result of changing the outcome in one match.  But then, due to \multNM{$2$}{$(3/2 - \varepsilon)$} it must be that the following three inequalities hold
    \begin{align*}
        r_i(T) + r_j(T) &> 2/3\\
        r_j(T) + r_k(T) &> 2/3\\
        r_k(T) + r_i(T) &> 2/3,
    \end{align*}
    since any of $i$, $j$, or $k$ could throw a match to produce a Condorcet winner.  On the other hand, we have that $r_i(T) + r_j(T) + r_k(T) = 1$ since $r$ must produce a probability distribution over the teams.  But then, adding the three inequalities together we obtain $2(r_i(T) + r_j(T) + r_k(T)) > 2$, which is a contradiction.
\end{proof}

\begin{figure}[h]
\centering

\tikzset{every picture/.style={line width=0.75pt}} 

\begin{tikzpicture}[x=0.75pt,y=0.75pt,yscale=-1,xscale=1]
\draw   (80,54) .. controls (80,40.19) and (91.19,29) .. (105,29) .. controls (118.81,29) and (130,40.19) .. (130,54) .. controls (130,67.81) and (118.81,79) .. (105,79) .. controls (91.19,79) and (80,67.81) .. (80,54) -- cycle ;
\draw   (197,54) .. controls (197,40.19) and (208.19,29) .. (222,29) .. controls (235.81,29) and (247,40.19) .. (247,54) .. controls (247,67.81) and (235.81,79) .. (222,79) .. controls (208.19,79) and (197,67.81) .. (197,54) -- cycle ;
\draw   (197,163) .. controls (197,149.19) and (208.19,138) .. (222,138) .. controls (235.81,138) and (247,149.19) .. (247,163) .. controls (247,176.81) and (235.81,188) .. (222,188) .. controls (208.19,188) and (197,176.81) .. (197,163) -- cycle ;
\draw   (80,163) .. controls (80,149.19) and (91.19,138) .. (105,138) .. controls (118.81,138) and (130,149.19) .. (130,163) .. controls (130,176.81) and (118.81,188) .. (105,188) .. controls (91.19,188) and (80,176.81) .. (80,163) -- cycle ;
\draw    (130,54) -- (195,54) ;
\draw [shift={(197,54)}, rotate = 180] [color={rgb, 255:red, 0; green, 0; blue, 0 }  ][line width=0.75]    (10.93,-3.29) .. controls (6.95,-1.4) and (3.31,-0.3) .. (0,0) .. controls (3.31,0.3) and (6.95,1.4) .. (10.93,3.29)   ;
\draw    (105,138) -- (105,81) ;
\draw [shift={(105,79)}, rotate = 90] [color={rgb, 255:red, 0; green, 0; blue, 0 }  ][line width=0.75]    (10.93,-3.29) .. controls (6.95,-1.4) and (3.31,-0.3) .. (0,0) .. controls (3.31,0.3) and (6.95,1.4) .. (10.93,3.29)   ;
\draw    (123,145.5) -- (203.5,74.82) ;
\draw [shift={(205,73.5)}, rotate = 138.72] [color={rgb, 255:red, 0; green, 0; blue, 0 }  ][line width=0.75]    (10.93,-3.29) .. controls (6.95,-1.4) and (3.31,-0.3) .. (0,0) .. controls (3.31,0.3) and (6.95,1.4) .. (10.93,3.29)   ;
\draw    (197,163) -- (132,163) ;
\draw [shift={(130,163)}, rotate = 360] [color={rgb, 255:red, 0; green, 0; blue, 0 }  ][line width=0.75]    (10.93,-3.29) .. controls (6.95,-1.4) and (3.31,-0.3) .. (0,0) .. controls (3.31,0.3) and (6.95,1.4) .. (10.93,3.29)   ;
\draw    (222,79) -- (222,136) ;
\draw [shift={(222,138)}, rotate = 270] [color={rgb, 255:red, 0; green, 0; blue, 0 }  ][line width=0.75]    (10.93,-3.29) .. controls (6.95,-1.4) and (3.31,-0.3) .. (0,0) .. controls (3.31,0.3) and (6.95,1.4) .. (10.93,3.29)   ;
\draw    (203,146.5) -- (122.5,75.82) ;
\draw [shift={(121,74.5)}, rotate = 41.28] [color={rgb, 255:red, 0; green, 0; blue, 0 }  ][line width=0.75]    (10.93,-3.29) .. controls (6.95,-1.4) and (3.31,-0.3) .. (0,0) .. controls (3.31,0.3) and (6.95,1.4) .. (10.93,3.29)   ;
\draw   (346,53) .. controls (346,39.19) and (357.19,28) .. (371,28) .. controls (384.81,28) and (396,39.19) .. (396,53) .. controls (396,66.81) and (384.81,78) .. (371,78) .. controls (357.19,78) and (346,66.81) .. (346,53) -- cycle ;
\draw   (463,53) .. controls (463,39.19) and (474.19,28) .. (488,28) .. controls (501.81,28) and (513,39.19) .. (513,53) .. controls (513,66.81) and (501.81,78) .. (488,78) .. controls (474.19,78) and (463,66.81) .. (463,53) -- cycle ;
\draw   (463,162) .. controls (463,148.19) and (474.19,137) .. (488,137) .. controls (501.81,137) and (513,148.19) .. (513,162) .. controls (513,175.81) and (501.81,187) .. (488,187) .. controls (474.19,187) and (463,175.81) .. (463,162) -- cycle ;
\draw   (346,162) .. controls (346,148.19) and (357.19,137) .. (371,137) .. controls (384.81,137) and (396,148.19) .. (396,162) .. controls (396,175.81) and (384.81,187) .. (371,187) .. controls (357.19,187) and (346,175.81) .. (346,162) -- cycle ;
\draw    (463,53) -- (398,53) ;
\draw [shift={(396,53)}, rotate = 360] [color={rgb, 255:red, 0; green, 0; blue, 0 }  ][line width=0.75]    (10.93,-3.29) .. controls (6.95,-1.4) and (3.31,-0.3) .. (0,0) .. controls (3.31,0.3) and (6.95,1.4) .. (10.93,3.29)   ;
\draw    (371,137) -- (371,80) ;
\draw [shift={(371,78)}, rotate = 90] [color={rgb, 255:red, 0; green, 0; blue, 0 }  ][line width=0.75]    (10.93,-3.29) .. controls (6.95,-1.4) and (3.31,-0.3) .. (0,0) .. controls (3.31,0.3) and (6.95,1.4) .. (10.93,3.29)   ;
\draw    (389,144.5) -- (469.5,73.82) ;
\draw [shift={(471,72.5)}, rotate = 138.72] [color={rgb, 255:red, 0; green, 0; blue, 0 }  ][line width=0.75]    (10.93,-3.29) .. controls (6.95,-1.4) and (3.31,-0.3) .. (0,0) .. controls (3.31,0.3) and (6.95,1.4) .. (10.93,3.29)   ;
\draw    (463,162) -- (398,162) ;
\draw [shift={(396,162)}, rotate = 360] [color={rgb, 255:red, 0; green, 0; blue, 0 }  ][line width=0.75]    (10.93,-3.29) .. controls (6.95,-1.4) and (3.31,-0.3) .. (0,0) .. controls (3.31,0.3) and (6.95,1.4) .. (10.93,3.29)   ;
\draw    (488,78) -- (488,135) ;
\draw [shift={(488,137)}, rotate = 270] [color={rgb, 255:red, 0; green, 0; blue, 0 }  ][line width=0.75]    (10.93,-3.29) .. controls (6.95,-1.4) and (3.31,-0.3) .. (0,0) .. controls (3.31,0.3) and (6.95,1.4) .. (10.93,3.29)   ;
\draw    (469,145.5) -- (388.5,74.82) ;
\draw [shift={(387,73.5)}, rotate = 41.28] [color={rgb, 255:red, 0; green, 0; blue, 0 }  ][line width=0.75]    (10.93,-3.29) .. controls (6.95,-1.4) and (3.31,-0.3) .. (0,0) .. controls (3.31,0.3) and (6.95,1.4) .. (10.93,3.29)   ;

\draw (99,49) node [anchor=north west][inner sep=0.75pt]   [align=left] {$a$};
\draw (217,46) node [anchor=north west][inner sep=0.75pt]   [align=left] {$b$};
\draw (218,159) node [anchor=north west][inner sep=0.75pt]   [align=left] {$c$};
\draw (99,156) node [anchor=north west][inner sep=0.75pt]   [align=left] {$d$};
\draw (365,49) node [anchor=north west][inner sep=0.75pt]   [align=left] {$a$};
\draw (483,45) node [anchor=north west][inner sep=0.75pt]   [align=left] {$b$};
\draw (484,159) node [anchor=north west][inner sep=0.75pt]   [align=left] {$c$};
\draw (365,155) node [anchor=north west][inner sep=0.75pt]   [align=left] {$d$};
\draw (157,209) node [anchor=north west][inner sep=0.75pt]   [align=left] {$T$};
\draw (433,209) node [anchor=north west][inner sep=0.75pt]   [align=left] {$T'$};

\end{tikzpicture}

\caption{Two $\{a,b\}$-adjacent tournaments $T$ and $T'$ demonstrating that \textsc{RandomizedSingleEliminationBracket} does not guarantee $2$-MNM-$\delta$ for any finite $\delta$.}
\label{fig:mnm-tournaments}

\end{figure}
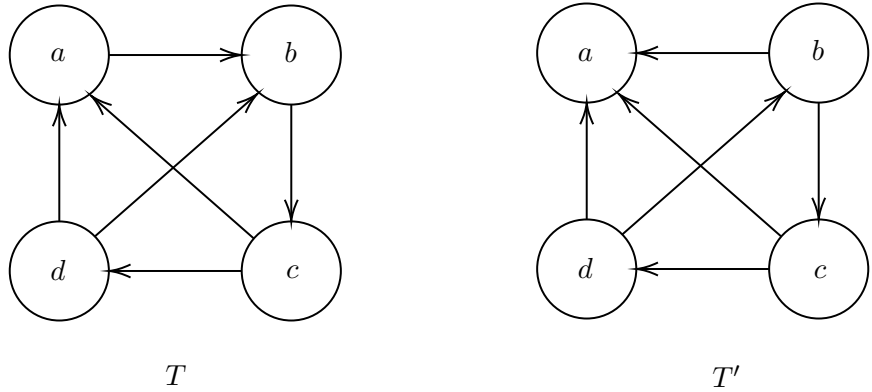

Given its connection and surface-level similarity to $2$-SNM-$\alpha$, one may hope that rules which guarantee $2$-SNM-$1/3$ (the best possible guarantee one can achieve \cite{schneider2017condorcet}) can also obtain \multNM{$2$}{$\delta$} for small values of $\delta$ (ideally even $\delta = 3/2$).  Unfortunately, this is not the case.  Indeed, as we demonstrate in \cref{obs:rseb-no-guar}, the \textsc{RandomizedSingleEliminationBracket} rule of \cite{schneider2017condorcet}, which guarantees Condorcet consistency, monotonicity, and $2$-SNM-$1/3$, fails to achieve \multNM{$2$}{$\delta$} for any bounded $\delta$.
\begin{definition}
    A \emph{single elimination bracket} for a tournament of $n = 2^h$ teams is a binary tree of height $h$.  Each leaf in the tree is labeled with one of the $n$ teams and each internal node is labeled with the team who wins the match between the its two child nodes.  The winner of the single elimination bracket is then the label of the root.  The \textsc{RandomizedSingleEliminationBracket} tournament rule for tournament $T$ assigns $r_i(T)$ to each team $i$ equal to the probability that $i$ is the winner of a uniformly random chosen single elimination bracket for $T$.
\end{definition}
\begin{observation}\label{obs:rseb-no-guar}
    The \textsc{RandomizedSingleEliminationBracket} tournament rule does not guarantee \multNM{$2$}{$\delta$} for any $\delta$.
\end{observation}
\begin{proof}
    Let $r$ denote the \textsc{RandomizedSingleEliminationBracket} tournament rule and consider the two tournaments $T$ and $T'$ in \Cref{fig:mnm-tournaments} over teams named $a$, $b$, $c$, and $d$.  Note that these tournaments are $\{a,b\}$-adjacent.  On one hand, observe that $a$ cannot win in any bracket for either $T$ and $T'$.  This is because $a$ is a Condorcet loser in $T'$ and $a$ must face one of $c$ or $d$ in any single elimination bracket over $T$ (either at height $0$ or at height $1$ if $a$ and $b$ are matched by the bracket at height $0$).  As a result, $r_a(T) = r_a(T') = 0$.  On the other hand, there is no bracket for $T$ where $b$ wins.  This is because if the bracket matches $b$ against $c$ at the leaves (the only possibility for which an internal node is labeled with $b$) then $b$ must be matched against $d$ at height $1$.  However, there is a bracket for $T'$ in which $b$ wins; namely, the bracket which matches $a$ and $b$ at the leaves allows $b$ to win the bracket.  But then, we have that $r_b(T) = 0$ and $r_b(T') >0$.  As such $r_a(T') + r_b(T') > \delta(r_a(T) + r_b(T))$ for any finite $\delta$.
\end{proof}

The counterexample for \textsc{RandomizedSingleEliminationBracket} provides some insight into the design of tournament rules achieving \multNM{$2$}{$\delta$}.  At minimum, if a tournament rule $r$ gives any of a pair of teams $i,j$ positive winning probability in one of two $\{i,j\}$-adjacent tournaments, it must also give the pair $i,j$ positive winning probability in the other adjacent tournament.  Importantly, the \emph{ratio} of winning probabilities for the teams should not deviate too much as a result of changing the outcome of one match.  With this observation in hand, we now can construct a tournament rule \multRule.

\begin{mybox}{\multRule}
\begin{enumerate}
    \item For each team $i \in N$ in tournament $T$, let $d_i$ denote the outdegree of vertex $i$ in the tournament graph of $T$ (i.e., the number of matches that $i$ wins).
    \item Compute the \emph{simple win score} $s_i$ of each team $i$ as $s_i = \left(\frac{1}{2}\right)^{(n-2) - d_i}$.
    \item If any team $i$ has $s_i = 2$, output $r_i(T) = 1$ and $r_j(T) = 0$ for all $j \neq i$.
    \item Otherwise, for each team $i \in N$, output \[r_i(T) = \frac{s_i}{\sum_{j \in N}{s_j}}.\]
\end{enumerate}
\end{mybox}

Our central result in this section is that the \multRule\ tournament rule achieves all of our desired properties.

\begin{theorem}\label{thm:mult-nm-main}
    The \multRule\ tournament rule is Condorcet consistent, monotone, and $3.5$-multiplicative $2$-non-manipulable for all tournaments.
\end{theorem}

\subsection{Proof of \Cref{thm:mult-nm-main}}

We first handle the easiest of the three properties, Condorcet consistency.  Essentially, \multRule\ satisfies Condorcet consistency by design.

\begin{observation}\label{obs:mult-cc}
    The \multRule\ tournament rule is Condorcet consistent. 
\end{observation}
\begin{proof}
    Observe that a Condorcet winner $i$ in tournament $T$ wins $n-1$ matches so $d_i = n-1$. Then, $i$'s simple win score $s_i = \left(\frac{1}{2}\right)^{(n-2) - (n-1)} = \left(\frac{1}{2}\right)^{-1} = 2$.  But then, $r_i(T) = 1$, as desired.
\end{proof}

We now turn to the more challenging properties of monotonicity and constant-multiplicative $2$-non-manipulability. 

\begin{lemma}\label{lem:mult-monotone}
    The \multRule\ tournament rule is monotone.
\end{lemma}
\begin{proof}
    Fix an arbitrary tournament $T$ and a pair of teams $i$ and $j$ with $i$ defeating $j$ in $T$. Let $T'$ denote the tournament which is $\{i,j\}$-adjacent to $T$.  We will show that $r_i(T) \geq r_i(T')$ (in fact, we will show $r_i(T) > r_i(T')$ and, hence, \multRule\ is strictly monotone).  Let $s_k(T)$ denote the simple score of team $k$ in $T$ and $s_k(T')$ denote the simple score of $k$ in $T'$.  Observe that for all $k \in N \setminus \{i,j\}$ we have $s_k(T) = s_k(T')$ (since the number of wins that each team other than $i$ and $j$ has is the same in $T$ and $T'$).  Since $i$ wins one fewer match in $T'$ than they do in $T$, we also have $s_i(T') = s_i(T)/2$.  Similarly, since $j$ wins one more match we have $s_i(T') = 2s_i(T)$.  As a result, we have 
    \begin{align*}
        r_i(T') &= \frac{s_i(T')}{\sum_{k \in N}s_k(T')}\\
        &= \frac{s_i(T')}{s_i(T') + s_j(T') + \sum_{k \in N \setminus \{i,j\}}{s_k(T')}}\\
        &= \frac{s_i(T)/2}{s_i(T)/2 + 2s_j(T) + \sum_{k \in N \setminus \{i,j\}}{s_k(T)}}\\
        &< \frac{s_i(T)/2}{s_i(T)/2 + \sum_{k \in N \setminus \{i\}}{s_k(T)}}\\
        &\leq \frac{s_i(T)}{\sum_{k \in N}{s_k(T)}}\\
        &= r_i(T),
    \end{align*}
    where the inequalities are due to the fact that $s_i(T)$ and $s_j(T)$ are both positive.
\end{proof}

We conclude the section by demonstrating that \multRule\ achieves \multNM{$2$}{$3.5$}.  At a very high level, \multRule\ achieves the guarantee for the following simple reasoning.  Changing the outcome of any match involving team $i$ only changes $s_i$ by a factor $2$ and does not affect the sum of simple win scores by too much.

\begin{lemma}\label{lem:mult-nm}
    The \multRule\ tournament rule satisfies $3.5$-multiplicative $2$-non-manipulability.
\end{lemma}
\begin{proof}
    Fix some tournament $T$ and some pair of teams $i$ and $j$ with $i$ defeating $j$ in $T$.  Let $T'$ denote the tournament which is $\{i,j\}$-adjacent to $T$.  We will show that $3.5(r_i(T) + r_j(T)) \geq r_i(T') + r_j(T')$.  We consider two cases based on the number of wins that $j$ has in $T$.

    \textbf{Case 1} \emph{$(j$ has fewer than $n-2$ wins in $T$)}\textbf{.}  For any team $k$ let $s_k(T)$ and $s_k(T')$ denote the simple win score of team $k$ in $T$ and $T'$, respectively.  As above, because $i$ wins one fewer game in $T'$ compared to $T$ and $j$ wins one more we have that $s_i(T') = s_i(T)/2$ and $s_j(T') = 2s_j(T)$.  Since in \multRule each team obtains winning probability proportional to their simple win score over the sum of simple win scores, we obtain \[r_i(T) + r_j(T) = \frac{s_i(T) + s_j(T)}{s_i(T) + s_j(T) + \sum_{k \in N \setminus \{i,j\}}{s_k(T)}},\]
    and 
    \begin{align*}
        r_i(T') + r_j(T') &= \frac{s_i(T') + s_j(T')}{s_i(T') + s_j(T') + \sum_{k \in N \setminus \{i,j\}}{s_k(T')}}\\
        &= \frac{s_i(T)/2 + 2s_j(T)}{s_i(T)/2 + 2s_j(T) + \sum_{k \in N \setminus \{i,j\}}{s_k(T)}}.
    \end{align*}
    But then, if $s_j(T) \geq s_i(T)$ we have that
    \begin{align*}
        r_i(T') + r_j(T') &= \frac{s_i(T)/2 + 2s_j(T)}{s_i(T)/2 + 2s_j(T) + \sum_{k \in N \setminus \{i,j\}}{s_k(T)}}\\
        &\leq \frac{s_i(T)/2 + 2s_j(T)}{3s_i(T)/2 + s_j(T) + \sum_{k \in N \setminus \{i,j\}}{s_k(T)}}\\
        &\leq \frac{2s_i(T) + 2s_j(T)}{s_i(T) + s_j(T) + \sum_{k \in N \setminus \{i,j\}}{s_k(T)}}\\
        &=2(r_i(T) + r_j(T)).
    \end{align*}
    Suppose instead that $s_j(T) < s_i(T)$.  Then, since the simple win sores form a geometric series with a common ratio $1/2$ we have that $2s_j(T) \leq s_i(T)$.  As such, we have \[2s_j(T) + \frac{s_i(T)}{2} \leq s_j(T) + \frac{s_i(T)}{2} + \frac{s_i(T)}{2} = s_i(T) + s_j(T).\]  On the other hand, we have that
    \[s_i(T) + s_j(T) + \sum_{k \in \setminus \{i,j\}}{s_k(T)} \leq 2\cdot \left(\frac{s_i(T)}{2} + 2s_j(T) + \sum_{k \in N \setminus \{i,j\}}{s_k(T)}\right).\]  Putting this altogether,
    \begin{align*}
        r_i(T') + r_j(T') &= \frac{s_i(T)/2 + 2s_j(T)}{s_i(T)/2 + 2s_j(T) + \sum_{k \in N \setminus \{i,j\}}{s_k(T)}}\\
        &\leq 2 \cdot \frac{s_i(T)/2 + 2s_j(T)}{s_i(T) + s_j(T) + \sum_{k \in N \setminus \{i,j\}}{s_k(T)}}\\
        &\leq \frac{2(s_i(T) + s_j(T))}{\sum_{k \in N \setminus \{i,j\}}{s_k}}\\
        &=2(r_i(T) + r_j(T)).
    \end{align*}
    In either case, $3.5(r_i(T) + r_j(T)) \geq 2(r_i(T) + r_j(T)) \geq r_i(T') + r_j(T')$, as desired.

    \textbf{Case 2} \emph{($j$ has $n-2$ wins in $T$)}\textbf{.}  Observe that if $i$ defeats $j$ in $T$ but $j$ has $n-2$ wins (i.e., team $j$ wins all other matches they participate in) then $j$ becomes a Condorcet winner in $T'$.  As such $r_i(T') + r_j(T') =1$.  It suffices then to show that $r_i(T) + r_j(T) \geq \frac{1}{3.5}$.  By definition, since $j$ defeated $n-2$ teams in $T$ we have that $s_j(T) = 1$.  As such, since $r_j(T) = s_j(T)/\sum_{k \in N}{s_k(T)}$ we just need to bound $\sum_{k \in N}{s_k(T)}$.  Consider that the possible values of $s_k$ for some $k$ as $d_k$ varies from $n-2$ down to $0$ is $1, \frac{1}{2}, \frac{1}{4}, \frac{1}{8}, \dots, 2^{-(n-2)}$.  Notice that this sequence is convex.  On the other hand, we have that if we index teams in any tournament in non-increasing order of wins then the partial sums for all $\ell \in [n]$ of the outdegrees in the tournament graph is  bounded as \[\sum_{i = 1}^{\ell}{d_i} \leq \sum_{i = 1}^{\ell}{(n-i)}.\]
    Since there is no Condorcet winner in $T$, we have that the outdegree maximizing $\sum_{k \in N}{s_k(T)}$ is $d_1 = d_2 = d_3 = (n-2)$ and $d_k=(n-k)$ for all $k > 3$.  Finally, this tells us that in $T$ we have that \begin{align*}
    \sum_{k \in N}{s_k(T)} \leq 1 + 1 + 1 + \sum_{\ell = 2}^{n}{\frac{1}{2^\ell}}
    \leq 3 + \sum_{\ell = 2}^{\infty}{\frac{1}{2^\ell}}
    &= 3.5.
    \end{align*}
    But then, $r_i(T') + r_j(T') \leq 3.5r_j(T)$, as desired.
\end{proof}

Combining \cref{obs:mult-cc} and \cref{lem:mult-monotone,lem:mult-nm}, we immediately obtain the proof of \cref{thm:mult-nm-main}

\section{Non-Manipulability for $\lambda$}\label{sec:lambda-main}
We now turn toward the more demanding notion of pairwise non-manipulability in the form of $2$-non-manipulability for $\lambda$. 
In building toward our rule giving $2$-NM$_{11}$, it is instructive to see where \multRule\ fails.  For this, we examine the ``true superman-kryptonite tournament'' on $n$ teams.

\begin{definition}
    A \emph{true superman-kryptonite} tournament $T$ on $n$ teams is the tournament in which teams can be indexed from $1$ to $n$ such that team $1$ (the ``superman'') who defeats all teams except team $n$ (the ``kryptonite'') and for all $i \neq 1$ and all $j > i$ team $i$ defeats team $j$.
\end{definition}

As Pennock et al. \cite{Pennock2024} demonstrated, the true superman-kryptonite construction acts as a strong counterexample for achieving $2$-NM$_\lambda$ for essentially all tournament rules in the literature.  It remains problematic for \multRule\ as well.  In particular, observe that as $n$ approaches infinity, \multRule\ assigns roughly $1/2$ winning probability to superman but only assigns $\left(\frac{1}{2}\right)^{n-3}$ probability to kryptonite.  As such, kryptonite can sacrifice exponentially small winning probability to allow superman to gain nearly $1/2$ winning probability (by making superman a Condorcet winner).

\begin{observation}
    The \multRule\ tournament rule does not guarantee $2$-NM$_\lambda$ for any $\lambda = o(n)$.
\end{observation}

Essentially, the issue lies in the fact that \multRule\ assigns winning probability to team $i$ based only on the \emph{number} of teams that $i$ beats, rather than the relative \emph{strength} of the teams that $i$ beats.  As such, the rule fails to adequately reward kryptonite in the true superman-kryptonite tournament.  With this insight in hand, we define the \lambdaRule\ tournament rule below, which, in effect, modifies \multRule\ to better capture team strength.

Note that the true win score of $i$ comprises both her own simple score and the sum of simple scores of teams $i$ is capable of beating (observe that her own simple score appears twice in her own true score), roughly capturing how many teams $i$ beats as well as the strength of those teams.  Before arguing about the properties of \lambdaRule, we first argue that it is a valid tournament rule, i.e., that it produces a probability distribution over the teams for any tournament (note that, unlike \multRule, we do not obtain this automatically).

\begin{mybox}{\multRule}
\begin{enumerate}
    \item For each team $i \in N$ in tournament $T$, let $d_i$ denote the number of matches that $i$ wins (i.e., the outdegree of vertex $i$ in the tournament graph of $T$) and let $D_i$ denote the set of teams $i$ defeats.
    \item Compute the \emph{simple win score} $s_i$ of each team $i$ as $s_i = \left(\frac{1}{2}\right)^{(n-2) - d_i}$.
    \item Compute the \emph{true win score} $t_i$ of each team $i$ as $t_i = 2s_i + \sum_{j \in D_i}{s_j}$
    \item If any team $i$ has $s_i = 2$, output $r_i(T) = 1$ and $r_j(T) = 0$ for all $j \neq i$.
    \item Otherwise, for each team $i \in N$, output \[r_i(T) = \frac{t_i}{14} + \frac{1}{n}\cdot\left(1 - \frac{\sum_{j \in N}{t_j}}{14}\right).\]
\end{enumerate}
\end{mybox}

\begin{lemma}\label{lem:lambda-prob-dist}
    For any tournament $T$ \lambdaRule\ produces a probability distribution over the teams, i.e., $\sum_{i \in N}{r_i(T)} = 1$.
\end{lemma}
\begin{proof}
    If $T$ has some team $i$ with $s_i = 2$ then \lambdaRule\ clearly outputs a probability distribution over the teams in $T$.  So suppose that no team has $s_i = 2$, i.e., each $s_i \leq 1$ and, hence, each $d_i \leq n-2$.  Then, letting $t_i(T)$ denote the true win score of of $i$ in $T$, it suffices to check that $\sum_{j \in N}{t_j(T)} \leq 14$.  Observe that by definition we have 
    \begin{align*}
        \sum_{i \in N}{t_i(T)} &= \sum_{i \in N}{\left(2s_i(T) + \sum_{j \in D_i}{s_j(T)}\right)}\\
        &=\sum_{i \in N}{((n-1) - d_i + 2)s_i}\\
        &=\sum_{i \in N}{\frac{(n-2)-d_i + 3}{2^{(n-2)-d_i}}},
    \end{align*}
    where the second equality comes from the fact that any team $i$ \emph{loses} to $(n-1) - d_i$ teams (and, thus, $s_i(T)$ is counted in the $t_j(T)$ of $(n-1) - d_i$ teams other than $i$).  Consider the sequence of values generated by $\frac{(n-2) - d_i + 3}{2^{(n-2)-d_i}}$ as we vary $d_i$ from $n-2$ down to $0$.  We have that this sequence is
    \[
    \frac{3}{1}, \frac{4}{2}, \frac{5}{4}, \frac{6}{8}, \dots, \frac{n}{2^{n-3}}, \frac{n+1}{2^{n-2}}.
    \]
    Observe that for all $k \geq 0$ we have that
    \[
    \frac{k+3}{2^k} + \frac{k+5}{2^{k+2}} = \frac{5k+17}{2^{k+2}} > \frac{4k+16}{2^{k+2}} = 2\cdot \frac{k+4}{2^{k+1}}.
    \]  But then, this sequence is convex.  As  before, we can then find the degree distribution maximizing $\sum_{j \in N}{t_j(T)}$.  We know that, for any tournament $T$, if we index the teams in non-increasing order of wins then we have that the partial sums of the outdegrees of the teams are bounded as $\sum_{i = 1}^{\ell}{d_i} \leq \sum_{i = 1}^{\ell}(n-i)$ for all $\ell$.  By convexity of the sequence, we then have that $\sum_{i =1}^{n}{t_i(T)}$ is maximized over all $T$ without a Condorcet winner by taking $d_1 = d_2 = d_3 = (n-2)$ and $d_k = n-k$ for all $k > 3$.
    But then, we may conclude by observing
    \begin{align*}
        \sum_{i \in N}{t_i(T)} &\leq 3 + 3 + \sum_{\ell=0}^{n-2}{\frac{(n-2) - \ell + 3}{2^{(n-2) - \ell}}}\\
        &\leq 6 + \sum_{\ell = 0}^{n-2}{\frac{\ell + 3}{2^\ell}}\\
        &\leq 6 + \sum_{\ell=0}^\infty{\frac{\ell + 3}{2^\ell}}\\
        &= 14,
    \end{align*}
    where the final equality follows from evaluating the sum of an infinite arithmetico-geometric sequence.
\end{proof}

\subsection{Properties of \lambdaRule}\label{sec:props-lambda}
We can now reason about the properties that our rule satisfies.  In particular, we argue that it satisfies Condorcet consistency, monotonicity, and $2$-NM$_{11}$.

Since any Condorcet winner $i$ has outdegree $n-1$ we have that $s_i = 2$ for Condorcet winner $i$.  As such, by Line 4 of \lambdaRule\ we obtain the Condorcet consistency guarantee.

\begin{observation}\label{obs:lambda-cc}
    The \lambdaRule\ tournament rule is Condorcet consistent.
\end{observation}

The monotonicity of \lambdaRule\ is less clear than that of \multRule, in part, since the $s_i$ is included in the true win score of \emph{all} teams that $i$ loses to.  As such, throwing a match may decrease $i$'s contribution to the true win scores of many teams.  In \cref{lem:lambda-monotone}, we show that, despite this, we achieve monotonicity.

\begin{lemma}\label{lem:lambda-monotone}
    The \lambdaRule\ tournament rule is monotone.
\end{lemma}
\begin{proof}
    Fix an arbitrary tournament $T$ and pair of teams $i$ and $j$ with $i$ defeating $j$.  Let $T'$ denote the tournament which is $\{i,j\}$-adjacent to $T$.  As before, we will show that $r_i(T) \geq r_i(T')$.  First, observe that if some other team $k \notin \{i,j\}$ is a Condorcet winner in $T$ then that team remains a Condorcet winner in $T'$ and $r_i(T) = r_i(T') = 0$.  So in the remainder of the proof, we will assume that there is no Condorcet winner in $T$.  We will consider two cases depending on how many wins $j$ has in $T$.

    \textbf{Case 1} \emph{($j$ has $n-2$ wins in $T$)}\textbf{.}  Let $s_i(T)$ and $t_i(T)$ denote the simple win score and true win score, respectively, of $i$ in $T$.  Observe that $s_i(T)$ is strictly positive so $t_i(T) > 0$.  But then, $t_i(T)/14 > 0$ and $r_i(T) > 0$.  On the other hand, since $j$ becomes a Condorcet winner in $T'$ we have that $r_i(T') = 0$.

    \textbf{Case 2} \emph{($j$ has fewer than $n-2$ wins in $T$)}\textbf{.}  We analyze the change in the two terms of $r_i$ as we move from $T$ to $T'$ separately.  Let $s_i(T)$ and $t_i(T)$ denote the simple win score and true win score, respectively of $i$ in $T$.

    First consider the change in $\frac{t_i}{14}$.  In moving from $T$ to $T'$ we have that $d_i$ decreases by $1$ and, hence, $s_i(T') = s_i(T)/2$.  Also, observe that $j \in D_i$ in $T$ but $j \notin D_i$ in $T'$.  As a result, we have that 
    \begin{equation}\label{eq:mono-phase1}
    t_i(T') = t_i(T) - (2(s_i(T)/2) + s_j(T)) = t_i(T) - (s_i(T) + s_j(T)).
    \end{equation}

    Now consider the the second term of $r_i$.  To compute the change in this term, we must analyze the change in total true win scores across all teams.  First, $j$ wins one more game in $T'$ than in $T$ so we have that $s_j(T') = 2s_j(T)$.  Let $c_i = |\{ k ~|~ i \in D_k, k \neq j \}|$ denote the number of teams other than $j$ that beat $i$ in $T$ (and also in $T'$) and let $c_j = |\{ k ~|~ j \in D_k, k \neq i \}|$ be defined similarly.  We then have that the total change in true win scores of teams other than $i$ and $j$ is equal to \[\sum_{k \in N \setminus \{i,j\}}{t_k(T') - t_k(T)} = - c_i\cdot s_i(T)/2 + c_j \cdot s_j(T).\] Altogether, we obtain 
    \begin{equation}\label{eq:mono-phase2}
        \sum_{k \in N}{t_k(T') - t_k(T)} = -(s_i(T) + s_j(T))+2s_j(T) + s_i(T)/2- c_i\cdot s_i(T)/2 + c_j \cdot s_j(T).
    \end{equation}
    Combining \cref{eq:mono-phase1,eq:mono-phase2} we have that
    \begin{align*}
        r_i(T') - r_i(T) = &-\frac{1}{14}(s_i(T) + s_j(T))\\&-\frac{1}{n} \cdot \frac{1}{14}\left(-(s_i(T) + s_j(T)) + 2s_j(T) + s_i(T)/2 + c_js_j(T) - c_is_i(T)/2\right),
    \end{align*}
    and hence
    \begin{align*}
        r_i(T') - r_i(T) &= -\frac{1}{14}(s_i(T) + s_j(T)) - \frac{1}{14n}((c_j + 1)s_j(T) - (c_i + 1)s_i(T)/2)
    \end{align*}
    But since $c_j \geq 0$ and $c_i \leq n-1$, we have
    \begin{align}\label{eq:prob-loss-i}
        r_i(T') - r_i(T) \leq -\frac{1}{14}(s_i(T)/2 + s_j(T)) \leq 0,
    \end{align}
    where the second inequality follows because $s_i(T)$ and $s_j(T)$ are non-negative.  Thus, $r_i(T) \geq r_i(T')$, as desired.
\end{proof}

We now turn toward showing that \lambdaRule\ satisfies $2$-NM$_{11}$. 

\begin{lemma}\label{lem:lambda-NM}
    The \lambdaRule\ tournament rule satisfies $2$-non-manipulability for $\lambda = 11$.
\end{lemma}
\begin{proof}
    Fix an arbitrary tournament $T$ and a pair of teams $i$ and $j$ with $i$ defeating $j$.  Let $T'$ denote the tournament which is $\{i,j\}$-adjacent to $T$.  Since \lambdaRule\ is monotone by \cref{lem:lambda-monotone}, we will show that the rule is $2$-NM$_{11}$ by demonstrating that $r_j(T') - r_j(T) \leq (11+1)(r_i(T) - r_i(T'))$.  As before, we proceed by case analysis.

    \textbf{Case 1} \emph{($j$ has $n-2$ wins in $T$)}\textbf{.}  Let $s_i(T)$ and $t_i(T)$ denote the simple win score and true win score, respectively, of team $i$ in tournament $T$.  Since $j$ defeats $n-2$ teams (i.e., all other teams besides $i$) in $T$, we have that $s_j(T) = 1$.  But then, since $j \in D_i$ in $T$ we have that $t_i(T) \geq 1$ and $t_j(T) \geq 2$ by definition of true win score.  This means, however, that $r_i(T) \geq 1/14$ and $r_j(T) \geq 2/14$.  In $T'$, since $j$ is a Condorcet winner, $r_j(T') = 1$ and $r_i(T') = 0$.  Then $r_j(T') - r_j(T) \leq 12/14 \leq (11 + 1)(r_i(T) -r_i(T'))$, as desired.

    \textbf{Case 2} \emph{($j$ has strictly fewer than $n-2$ wins in $T$)}\textbf{.}  Once again, let $s_i(T)$ and $t_i(T)$ denote the simple win score and true win score, respectively of team $i$ in tournament $T$. This case then follows similar analysis to the analogous case in \cref{lem:lambda-monotone}.  Recall from \cref{eq:prob-loss-i} that going from $T$ to $T'$ yields a change in winning probability for team $i$ at most
    \begin{equation}\label{eq:i-decrease}
    r_i(T) - r_i(T') \leq -\frac{1}{14}\left(\frac{s_i(T)}{2} + s_j(T)\right).
    \end{equation}
    We then need to compute $r_j(T) - r_j(T')$.  First observe that since $j$ wins one more match in $T'$ compared to $T$ and since $i \in D_j$ in $T'$ we have that $s_j(T') = 2s_j(T)$ and, as a result,  \[t_j(T') - t_j(T) = 2s_j(T) + s_i(T)/2.\]  As in the proof of \cref{lem:lambda-monotone}, defining $c_i = |\{ k ~|~ i \in D_k, k \neq j \}|$ to be the number of teams other than $j$ that beat $i$ in $T$ (and also in $T'$) and defining $c_j = |\{ k ~|~ j \in D_k, k \neq i \}|$ analogously we have from \cref{eq:mono-phase2} that \begin{align*}
    \sum_{k\in N}{t_k(T') - t_k(T)} &= -(s_i(T) + s_j(T))+2s_j(T) + s_i(T)/2- c_i\cdot s_i(T)/2 + c_j \cdot s_j(T)\\
    &=(c_j+1)\cdot s_j(T) - (c_i+1)\cdot s_i(T)/2
    \end{align*}
    Putting this all together, we obtain that the change in $j$'s winning probability is 
    \begin{align}\label{eq:j-increase}
        r_j(T') - r_j(T) &= \frac{1}{14} \cdot (2s_j(T) + s_i(T)/2) - \frac{1}{14n}((c_j+1)\cdot s_j(T) - (c_i+1)\cdot s_i(T)/2)\nonumber\\
        &\leq \frac{1}{14}\cdot(2s_j(T) + s_i(T)),
    \end{align}
    where the last line comes from the fact that $c_j \geq 0$ and $c_i \leq n-1$.

    But then, combining \cref{eq:i-decrease,eq:j-increase}, we have that the increase in $j$'s winning probability is at most $2$ times the decrease in $i$'s winning probability, satisfying $2$-NM$_{11}$ in this case, as desired. 
\end{proof}

Combining \cref{lem:lambda-prob-dist,lem:lambda-monotone,lem:lambda-NM}, and \cref{obs:lambda-cc}, we obtain the following.

\begin{theorem}
    The \lambdaRule\ tournament rule satisfies Condorcet consistency, monotonicity, and $2$-non-manipulability for $\lambda = 11$ for all tournaments with any number of teams.
\end{theorem}

\section{Discussion and Conclusions}
In this work, we make a major step toward understanding the design of tournament rules for teams with a constant degree of selfishness.  We design a tournament rule satisfying monotonicity, Condorcet consistency, and $2$-non-manipulability for $\lambda = 11$.  Notably, this is the first tournament rule which satisfies Condorcet consistency and $2$-NM$_\lambda$ for any $\lambda = o(n)$.  We also introduce the notion of
$\delta$-multiplicative $k$-non-manipulability and provide a rule achieving monotonicity, Condorcet consistency, and \multNM{$2$}{$3.5$}.  In Appendix \Cref{sec:tcc-mult-nm}, we explore fairness rules strengthening Condorcet consistency.  In particular, we consider \emph{top cycle consistency} (TCC) and provide a rule satisfying TCC, monotonicity, and \multNM{$2$}{$5$}.

Many interesting questions arise from our work.  First, although we make strides toward closing the gap identified by \cite{Pennock2024}, determining whether there exists a Condorcet consistent tournament rule satisfying $2$-NM$_1$ is a compelling open question.  Relatedly, determining the smallest possible $\delta$ that a Condorcet consistent rule can achieve in terms of $k$-MNM-$\delta$ (for all $k$) is another potential fruitful line of work.  Finally, determining whether there exists a tournament rule satisfying stronger fairness notions, e.g., top cycle consistency or cover consistency (a further refinement), alongside $2$-NM$_{O(1)}$ is an important line of extension.

\section*{Acknowledgments}
This work was supported by a grant to DIMACS from the Simons Foundation (820931).

\bibliographystyle{splncs04}
\bibliography{bibliography}

@inproceedings{altman2010nonmanipulable2,
  title={Nonmanipulable randomized tournament selections},
  author={Altman, Alon and Kleinberg, Robert},
  booktitle={Proceedings of the AAAI Conference on Artificial Intelligence},
  volume={24},
  number={1},
  pages={686--690},
  year={2010}
}

@inproceedings{altman2010nonmanipulable1,
  title={Nonmanipulable selections from a tournament},
  author={Altman, Alon and Procaccia, Ariel D and Tennenholtz, Moshe},
  booktitle={Dagstuhl Seminar Proceedings},
  year={2010},
  organization={Schloss Dagstuhl-Leibniz-Zentrum f{\"u}r Informatik}
}

@inproceedings{Pennock2024,
author = {Pennock, David and Schvartzman, Ariel and Xue, Eric},
title = {Toward Fair and Strategyproof Tournament Rules for Tournaments with Partially Transferable Utilities},
year = {2024},
isbn = {978-3-031-73902-6},
publisher = {Springer-Verlag},
address = {Berlin, Heidelberg},
booktitle = {Algorithmic Decision Theory: 8th International Conference, ADT 2024, New Brunswick, NJ, USA, October 14–16, 2024, Proceedings},
pages = {174–188},
numpages = {15},
location = {New Brunswick, NJ, USA}
}

@inproceedings{schneider2017condorcet,
  title={Condorcet-Consistent and Approximately Strategyproof Tournament Rules},
  author={Schneider, Jon and Schvartzman, Ariel and Weinberg, S Matthew},
  booktitle={8th Innovations in Theoretical Computer Science Conference (ITCS 2017)},
  year={2017},
  organization={Schloss Dagstuhl--Leibniz-Zentrum f{\"u}r Informatik}
}

@inproceedings{schvartzman2020approximately,
  title={Approximately strategyproof tournament rules: On large manipulating sets and cover-consistence},
  author={Schvartzman, Ariel and Weinberg, S Matthew and Zlatin, Eitan and Zuo, Albert},
  booktitle={11th Innovations in Theoretical Computer Science Conference, ITCS 2020},
  pages={3},
  year={2020},
  organization={Schloss Dagstuhl-Leibniz-Zentrum fur Informatik GmbH, Dagstuhl Publishing}
}

@inproceedings{ding2021approximately,
  title={Approximately Strategyproof Tournament Rules in the Probabilistic Setting},
  author={Ding, Kimberly and Weinberg, S Matthew},
  booktitle={12th Innovations in Theoretical Computer Science Conference (ITCS 2021)},
  year={2021},
  organization={Schloss Dagstuhl-Leibniz-Zentrum f{\"u}r Informatik}
}

@inproceedings{dinev2022,
author = {Dinev, Atanas and Weinberg, S. Matthew},
title = {Tight Bounds on 3-Team Manipulations in Randomized Death Match},
year = {2022},
isbn = {978-3-031-22831-5},
publisher = {Springer-Verlag},
address = {Berlin, Heidelberg},
booktitle = {Web and Internet Economics: 18th International Conference, WINE 2022, Troy, NY, USA, December  12–15, 2022, Proceedings},
pages = {273–291},
numpages = {19},
location = {Troy, NY, USA}
}

@inproceedings{miksanik2024,
author = {Mik\v{s}an\'{\i}k, David and Schvartzman, Ariel and Soukup, Jan},
title = {On Approximately Strategy-Proof Tournament Rules for Collusions of Size at Least Three},
year = {2024},
isbn = {978-3-031-73902-6},
publisher = {Springer-Verlag},
address = {Berlin, Heidelberg},
booktitle = {Algorithmic Decision Theory: 8th International Conference, ADT 2024, New Brunswick, NJ, USA, October 14–16, 2024, Proceedings},
pages = {33–47},
numpages = {15},
location = {New Brunswick, NJ, USA}
}

@inproceedings{suksompong2021tournaments,
  title={Tournaments in Computational Social Choice: Recent Developments.},
  author={Suksompong, Warut},
  booktitle={IJCAI},
  pages={4611--4618},
  year={2021}
}

@article{aziz2018fixing,
  title={Fixing balanced knockout and double elimination tournaments},
  author={Aziz, Haris and Gaspers, Serge and Mackenzie, Simon and Mattei, Nicholas and Stursberg, Paul and Walsh, Toby},
  journal={Artificial Intelligence},
  volume={262},
  pages={1--14},
  year={2018},
  publisher={Elsevier}
}

@inproceedings{kim2015fixing,
  title={Fixing tournaments for kings, chokers, and more},
  author={Kim, Michael P and Vassilevska Williams, Virginia},
  booktitle={Twenty-Fourth International Joint Conference on Artificial Intelligence},
  year={2015}
}

@book{laslier1997tournament,
  title={Tournament solutions and majority voting},
  author={Laslier, Jean-Fran{\c{c}}ois},
  volume={7},
  year={1997},
  publisher={Springer}
}

@book{brandt2016handbook,
  title={Handbook of computational social choice},
  author={Brandt, Felix and Conitzer, Vincent and Endriss, Ulle and Lang, J{\'e}r{\^o}me and Procaccia, Ariel D},
  year={2016},
  publisher={Cambridge University Press}
}

@article{shpigel2023important,
  title={This is why important World Cup games are played at the same time.},
  author={Shpigel, Ben},
  journal={The New York Times (Digital Edition)},
  pages={NA--NA},
  year={2023},
  publisher={The New York Times Company}
}

@article{Lang2012,
author = {Lang, J\'{e}r\^{o}me and Pini, Maria Silvia and Rossi, Francesca and Salvagnin, Domenico and Venable, Kristen Brent and Walsh, Toby},
title = {Winner determination in voting trees with incomplete preferences and weighted votes},
year = {2012},
issue_date = {July      2012},
publisher = {Kluwer Academic Publishers},
address = {USA},
volume = {25},
number = {1},
issn = {1387-2532},
journal = {Autonomous Agents and Multi-Agent Systems},
month = jul,
pages = {130–157},
numpages = {28}
}

@book{moulin1991axioms,
  title={Axioms of cooperative decision making},
  author={Moulin, Herv{\'e}},
  number={15},
  year={1991},
  publisher={Cambridge university press}
}

\appendix
\section{Refined Fairness Guarantees}\label{sec:tcc-mult-nm}
In this section, we consider whether it is possible to provide some stronger guarantee on fairness than Condorcet consistency.  Indeed, while both \multRule\ and \lambdaRule\ guarantee that a Condorcet winner $i$ in tournament $T$ gets $r_i(T) = 1$, they both give a \emph{Condorcet loser} (i.e., some team which loses all pairwise matches) a (small) positive probability of winning.  This is because the smallest simple win score that either rule assigns to any player is $\frac{1}{2^{n-2}} > 0$.  We would like, then, to refine the notion of Condorcet consistency to obtain better tournament fairness (and, hence, better tournament competitiveness).  For this, we seek a tournament rule guaranteeing \emph{top cycle consistency} (abbreviated TCC). 

\begin{definition}
    The \emph{top cycle} of a tournament $T$ is the (inclusion-wise) minimal subset $S \subseteq N$ of teams such that for all $i \in S$ and all $j \in N \setminus S$ we have that $i$ defeats $j$.  The top cycle of a tournament is always non-empty and unique.  A tournament rule $r$ satisfies \emph{top cycle consistency} if for all $i \in N$ if $r_i(T) > 0$ then $i$ is in the top cycle of $T$.
\end{definition}

Observe that TCC is a refinement of Condorcet consistency since if a tournament contains a Condorcet winner, then the top cycle comprises exactly the Condorcet winner.  As argued above, while both of our previous tournament rules satisfy Condorcet consistency, neither satisfies TCC (since a Condorcet loser is, by definition, not in the top cycle).  We can, however, modify \multRule\ in a straightforward way to obtain a tournament rule satisfying TCC, monotonicity, and \multNM{$2$}{$\delta$}.  At a high level, we can achieve TCC by simply ``zeroing out'' the contributions of any team not contained in the top cycle.  We formalize this in the following rule.

\begin{mybox}{\multTCCRule}
\begin{enumerate}
    \item Compute the top cycle $S_T$ of $T$ and let $n' = |S_T|$.\footnotemark
    \item For each team $i \in S_T$ in tournament $T$, let $d_i$ be the number of wins that team $i$ has \emph{only counting teams in $S_T$}.
    \item Compute the simple win score $s_i$ of each team $i\in S_T$ as $s_i = \left(\frac{1}{2}\right)^{(n'-2) - d_i}$ and for each team $i \in N \setminus S_T$ let $s_i = 0$.
    \item For each team $i \in N$, output \[r_i(T) = \frac{s_i}{\sum_{j \in N}{s_j}} = \frac{s_i}{\sum_{j \in S_T}{s_j}}.\]
\end{enumerate}
\end{mybox}
\footnotetext{This step does not introduce any computational complexity concerns since the top cycle of a tournament can be computed by a greedy-like algorithm in time polynomial in the number of teams \cite{Lang2012}.}

\subsection{Properties of \multTCCRule}

\begin{observation}\label{obs:mult-tcc}
    The \multTCCRule\ tournament rule is TCC.
\end{observation}
\begin{proof}
    Observe that any team $i$ outside of the top cycle of $T$ has their simple win score $s_i = 0$.  Since the top cycle is always non-empty and unique, we also have that $\sum_{j \in S}{s_j} > 0$ in $T$.  As such, $r_i(T) = 0$.
\end{proof}

The proofs that the \multTCCRule\ tournament rule is monotone and satisfies \multNM{$2$}{$5$} follow a similar pattern to the proofs of \cref{lem:mult-monotone,lem:mult-nm}, respectively, but we need to treat carefully the cases when two adjacent tournament have different top cycles.  We begin by showing that \multTCCRule\ satisfies monotonicity.

\begin{lemma}\label{lem:tcc-mono}
The \multTCCRule\ tournament rule is monotone.
\end{lemma}
\begin{proof}
    Fix an arbitrary tournament $T$ and pair of teams $i$ and $j$ with $i$ defeating $j$ in $T$. Let $T'$ denote the tournament which is $\{i,j\}$-adjacent to $T$.  We will show that $r_i(T) \geq r_i(T')$, as needed.  Let $s_k(T)$ and $s_k(T')$ denote the simple scores of team $k$ in $T$ and $T'$, respectively.
    
    First consider the case that $i$ is not in the top cycle of $T$, which is the simplest.  Since $i$ is not in the top cycle of $T$, we have, by top cycle consistency of \multTCCRule, that $r_i(T) = 0$.  But if $i$ is not in the top cycle of $T$ then $i$ is also not in the top cycle of $T'$ (since losing an additional game cannot move $i$ into the top cycle).  But then $r_i(T) = r_i(T') = 0$.

    Now consider the case where both $i$ and $j$ are in the top cycle in $T$.  This case turns out to be very similar to the proof of \cref{lem:mult-monotone}.  First, if $i$ is not in the top cycle of $T'$ then $r_i(T) > r_i(T')$ since $r_i(T) > 0$ (since $s_k(T) > 0$ for any team $k \in N$ in the top cycle of $T$) and $r_i(T') = 0$ (since $s_k(T') = 0$ for any team $k \in N$ not in the top cycle of $T'$).  So suppose that $i$ remains in the top cycle of $T'$.  Let $S_T$ denote the top cycle of $T$ and $S_{T'}$ denote the top cycle of $T'$.  Note that since $T$ and $T'$ are $\{i,j\}$-adjacent, we must have that $S_T = S_{T'}$.  This is because a team $k$ is in the top cycle if and only if there is a path in the tournament graph from $k$ to each other team $k'$ (see, e.g., \cite{moulin1991axioms}), and hence, no teams leave nor enter the top cycle in moving from $T$ to $T'$.  Then,  for all $k \in N \setminus \{i,j\}$ we have that $s_k(T) = s_k(T')$ and for $i$ and $j$ we obtain that $s_i(T') = s_i(T)/2$ and $s_j(T') = 2s_j(T)$.  As a result, by following the exact same (in)equalities as in the proof of \cref{lem:mult-monotone}, we obtain that $r_i(T) \geq r_i(T')$.

    We now consider the final case where $i$ is in the top cycle of $T$ and $j$ is not.  In this case, moving to $T'$ can change the top cycle.  As before, if $i$ is not in the top cycle of $T'$ then we have that $r_i(T) \geq r_i(T')$ by design.  So suppose that $i$ remains in the top cycle of $T'$.  By definition, it must be that $j$ is in the top cycle of $T'$.  Letting $S_T$ denote the top cycle of $T$ and $S_{T'}$ denote the top cycle of $T'$, we then have that $S_T \subsetneq S_{T'}$ since any team in $S_T$ remains in $S_{T'}$ as a result of these teams beating $j$.  But then, we may conclude that $s_i(T') = s_i(T)/2$ (since $i$ wins one fewer game but $i \in S_{T'}$), $s_j(T') > 2s_j(T)$ (since $s_j(T) = 0$ whereas $s_j(T')$ is positive) and $s_k(T') \geq s_k(T)$ for all $k \in N \setminus \{i,j\}$ (since all teams in $S_T \cap S_{T'}$ have the same simple win score in both tournaments and all teams in $S_{T'} \setminus S_T$ have positive simple win score in $T'$).  But then, once more, we may follow the same sequence of (in)equalities as in the proof of \cref{lem:mult-monotone} to conclude that $r_i(T) \geq r_i(T')$.
\end{proof}

We now turn toward the the pairwise non-manipulability of \multTCCRule.

\begin{lemma}\label{lem:tcc-nm}
    \multTCCRule\ satisfies $5$-multiplicative $2$-non-manipulability.
\end{lemma}
\begin{proof}
    Fix some tournament $T$ and some pair of teams $i$ and $j$ with $i$ defeating $j$ in $T$.  Let $T'$ denote the tournament which is $\{i,j\}$ adjacent to $T$.  We will show $r_i(T') + r_j(T') \leq 5 (r_i(T) + r_j(T))$. Let $s_k(T)$ and $s_k(T')$ denote the simple win scores of team $k$ in $T$ and $T'$, respectively.

    As before, first consider the case that $i$ is not in the top cycle of $T$.  Then, $j$ is not in the top cycle of $T$, but also $j$ is not in the top cycle of $T'$ (since $i$ does not move into the top cycle in $T'$).  As a result, $r_i(T) + r_j(T) = r_i(T') + r_j(T') = 0$ by the top cycle consistency of \multTCCRule.

    Now consider the case where $i$ is in the top cycle of $T$ but $j$ is not. 
    
    Suppose first that $i$ is not in the top cycle of $T'$ but $j$ is in the top cycle of $T'$.  Then it must be that $i$ is a Condorcet winner in $T$ and $j$ is a Condorcet winner in $T'$.  This is because $T$ and $T'$ are $\{i,j\}$-adjacent and $j$ is not in the top cycle of $T$ so any team $k \neq i$ in the top cycle of $T$ must remain in the top cycle of $T'$ and, as a result, $k$ pairwise beats the same set of teams across $T$ and $T'$.  But then, each $k\neq i$ in the top cycle of $T$ must also beat $j$ in $T'$.  As a result, since the top cycle is minimal inclusion-wise, if any $k \neq i$ is in the top cycle of $T$ then $j$ cannot be in the top cycle of $T'$.  But then, by top cycle consistency of \multTCCRule, if $i$ is not in the top cycle of $T'$ but $j$ is, then we have that $r_i(T) = r_j(T') = 1$.  
    
    Suppose instead that both $i$ and $j$ are in the top cycle of $T'$.  Letting $S_T$ and $S_{T'}$ denote the top cycle of $T$ and $T'$, respectively, we then have that $S_T \subsetneq S_{T'}$.  On one hand, for all $k \in S_{T'}\setminus S_T$ with $k \neq j$ we have that $s_k(T') > s_k(T)$ and for all $k \notin S_{T'}$ we have $s_k(T') = s_k(T) = 0$.  On the other, for any team $k \in S_{T}$ with $k\neq i$ we have $s_k(T') = s_k(T)$ (since all teams in $S_T$ other than $i$ beat all teams entering the top cycle).  Putting this all together, we obtain
    $\sum_{k \in N \setminus \{i,j\}}s_k(T') \geq \sum_{k \in N \setminus \{i,j\}}s_k(T).$
    It thus remains to consider the effect of the change on the simple win scores of $i$ and $j$.  Observing that in $T'$ $i$ defeats all teams in $S_{T'} \setminus S_T$ except for $j$, we have that $s_i(T') = s_i(T)/2$.  However, we also know that $i$ must defeat every team that $j$ defeats in $T'$ (since $i$ was in the top cycle of $T$).  Indeed, there must exist some $k \in S_T$ such that $i$ defeats $k$ in $T$ (as otherwise $i$ would not be in $S_T$), but $j$ cannot beat this $k$ (since $j$ was not in $S_T$).  As a result, $i$ beats at least one team in $S_{T'}$ that $j$ does not beat.  This means that $s_j(T') \leq s_i(T') = s_i(T)/2$.  But then, we have that
    \begin{align*}
        r_i(T') + r_j(T') &= \frac{s_i(T') + s_j(T')}{s_i(T') + s_j(T') + \sum_{k\in N \setminus \{i,j\}}{s_k(T')}}\\
        &\leq \frac{s_i(T)/2 + s_i(T)/2}{s_i(T)/2 + \sum_{k\in N \setminus \{i,j\}}{s_k(T)}}\\
        &\leq \frac{s_i(T)}{s_i(T) + \sum_{k\in N \setminus \{i,j\}}{s_k(T)}}\\
        &=\frac{s_i(T) + s_j(T)}{\sum_{k\in N}{s_k(T)}}\\
        &= r_i(T) + r_j(T),
    \end{align*}
    where the second equality comes from the fact that $s_j(T) = 0$ since $j \notin S_T$.

    Finally, consider the most subtle case where both $i$ and $j$ are in the top cycle of $T$.  Once more, let $S_T$ and $S_{T'}$ denote the top cycles of $T$ and $T'$.  
    
    Suppose first that $i \in S_{T'}$ (i.e., $i$ remains in the top cycle after the change).  Then, $S_T = S_{T'}$.  As a result, there is little change in the overall structure and we have that $s_i(T') = s_i(T)/2$ while $s_j(T') = 2s_j(T)$.  In this situation, we can proceed essentially identically to the proof of \cref{lem:mult-nm} and consider the relative value of $s_i(T)$ and $s_j(T)$.  Observe that if $s_j(T) \geq s_i(T)$ then 
    \begin{align*}
        r_i(T') + r_j(T') &= \frac{s_i(T)/2 + 2s_j(T)}{s_i(T)/2 + 2s_j(T) + \sum_{k \in N \setminus \{i,j\}}{s_k(T)}}
        \leq \frac{2s_i(T) + 2s_j(T)}{\sum_{k \in N}{s_k(T)}}
        =2(r_i(T) + r_j(T)).
    \end{align*}
    On the other hand, if $s_j(T) < s_i(T)$ then we have $2s_j(T) \leq s_i(T)$.  As such, we have \[2s_j(T) + \frac{s_i(T)}{2} \leq s_j(T) + \frac{s_i(T)}{2} + \frac{s_i(T)}{2} = s_i(T) + s_j(T).\]  We also have that
    \[s_i(T) + s_j(T) + \sum_{k \in \setminus \{i,j\}}{s_k(T)} \leq 2\cdot \left(\frac{s_i(T)}{2} + 2s_j(T) + \sum_{k \in N \setminus \{i,j\}}{s_k(T)}\right).\]  Putting this altogether,
    \begin{align*}
        r_i(T') + r_j(T') &= \frac{s_i(T)/2 + 2s_j(T)}{s_i(T)/2 + 2s_j(T) + \sum_{k \in N \setminus \{i,j\}}{s_k(T)}}\\
        &\leq 2 \cdot \frac{s_i(T)/2 + 2s_j(T)}{s_i(T) + s_j(T) + \sum_{k \in N \setminus \{i,j\}}{s_k(T)}}\\
        &\leq \frac{2(s_i(T) + s_j(T))}{\sum_{k \in N}{s_k}}\\
        &=2(r_i(T) + r_j(T)).
    \end{align*}

    Now suppose that $i \notin S_{T'}$ (i.e., $i$ leaves the top cycle after the change).  When $i$ is no longer in the top cycle, this can cause other teams in $S_T$ to not be in $S_{T'}$.  However, for any team $k \in S_T \setminus S_{T'}$ we may make the following crucial observation, namely, that $j$ must defeat $k$ in $T$.  More strongly, for any $k \in S_T \setminus S_{T'}$ the only teams $k$ defeats in $S_T$ in tournament $T$  must be contained in the set $S_T \setminus S_{T'}$.  Moreover, any team $k' \in S_{T'}$ must beat each $k \in S_T \setminus S_{T'}$ in $T'$ (by definition).  As a result, we immediately obtain that $s_k(T') = s_k(T)$ for all $k \in S_{T'} \setminus \{j\}$ and for all $k \notin S_T$ (since no teams enter the top cycle as a result of $i$ leaving).  We next would like to bound the change in simple win scores of teams which are in $S_T \setminus S_{T'}$.  Let $m = |S_T \setminus S_{T'}|$.  Since any team in $S_T \setminus S_{T'}$ only defeats other teams in $S_T \setminus S_{T'}$ there are at most $m \choose 2$ matches between these teams.  Examining the induced tournament subgraph on the $m$ vertices corresponding to $S_T \setminus S_{T'}$ and indexing these vertices in non-increasing order of wins, we have that the partial sums of the outdegrees is bounded for all $\ell \in [m]$ as $\sum_{i = 1}^{\ell}{m-i}$.  But then, by convexity of the sequence $\left(\frac{1}{2}\right)^{(n' - 2) - d_i}$ in $d_i$, we have that the sum of the simple win scores of the teams in $S_T \setminus S_{T'}$ is maximized when the $\ell$-th team in non-increasing win order has $m-\ell$ wins among teams in $S_T \setminus S_{T'}$.  As a result, we have that \[\sum_{k \in S_T \setminus S_{T'}}{s_k(T)} \leq \sum_{k = 1}^{m}{\left(\frac{1}{2}\right)^{(n'-2) - k}} \leq \left(\frac{1}{2}\right)^{(n'-2) - (m+1)} \leq 4s_j(T),\]
    where the second inequality comes from evaluating the geometric sequence and the final inequality comes from the fact that $j$ defeats all teams in $S_T \setminus S_{T'}$ except for $i$ in $T$ (and, hence, $j$ has at least $m-1$ wins).
    Putting this all together, we may observe that $s_j(T') = 2s_j(T)$ since the size of the top cycle decreases by $1$ more than the number of $j$'s wins in the top cycle (due to $j$ beating $i$ in $T'$).  This allows us to conclude that 
    \begin{align*}
        r_i(T') + r_j(T') &= \frac{s_j(T')}{\sum_{k \in S_{T'}}{s_k(T')}}\\
        &=\frac{s_j(T')}{s_j(T') + \sum_{k \in S_{T'} \setminus \{j\}}{s_k(T')}}\\
        &=\frac{2s_j(T)}{2s_j(T) + \sum_{k \in S_{T'} \setminus \{j\}}{s_k(T)}}\\
        &=\frac{5s_j(T)}{s_j(T) + 4s_j(T) + 2.5 \sum_{k \in S_{T'} \setminus \{j\}}{s_k(T)}}\\
        &\leq \frac{5s_j(T)}{s_j(T) + \sum_{k \in S_T \setminus S_{T'}}{s_k(T)} + \sum_{k \in S_{T'}\setminus \{j\}}{s_k(T)}}\\
        &\leq 5(r_i(T) + r_j(T)),
    \end{align*}
    as desired, completing the proof.
\end{proof}

Combining \cref{obs:mult-tcc} with \cref{lem:tcc-mono,lem:tcc-nm}, we obtain the following theorem.

\begin{theorem}
        The \multTCCRule\ tournament rule is top cycle consistent, monotone, and $5$-multiplicative $2$-non-manipulable for all tournaments with any number of teams.
\end{theorem}

Unfortunately, a similar straightforward approach of ``pruning'' out the teams outside of the top cycle does not seem to work for \lambdaRule.  This is because teams are not declared winners directly in proportion to their true win score. Rather, we need to divide each $t_i$ by an upper bound on the sum of scores. Designing a tournament rule satisfying top cycle consistency, monotonicity, and $2$-NM$_\lambda$ for $\lambda = O(1)$ is an interesting and seemingly challenging open question.

\end{document}